\documentstyle[11pt]{article} 

\newcommand{\lood}{^\perp}

\newcommand{\1}{_{1}}
\newcommand{\2}{_{2}}

\newcommand{\en}{\&}
\newcommand{\of}{\vee}

\newcommand{\volgt}{\vdash}

\newtheorem{th}{Theorem}  
\newtheorem{ax}{Axiom}  
\newtheorem{lm}{Lemma} 
\newtheorem{df}{Definition}    
\newtheorem{pr}{Proposition} 
\newtheorem{cl}{Corollary}  
\newtheorem{re}{Remark}    
\newtheorem{as}{Assumption}  
\newtheorem{wg}{Wild Guess}
\newtheorem{ex}{Example}
\newcommand{\bth}{\begin{th}\hspace{-5pt}{\bf .} \ } 
\newcommand{\eth}{\end{th}}
\newcommand{\bax}{\begin{ax}\hspace{-5pt}{\bf .} \ } 
\newcommand{\eax}{\end{ax}}
\newcommand{\blm}{\begin{lm}\hspace{-5pt}{\bf .} \ }
\newcommand{\elm}{\end{lm}}
\newcommand{\bdf}{\begin{df}\hspace{-5pt}{\bf .} \ }   
\newcommand{\edf}{\end{df}} 
\newcommand{\bpr}{\begin{pr}\hspace{-5pt}{\bf .} \ } 
\newcommand{\epr}{\end{pr}}
\newcommand{\bcl}{\begin{cl}\hspace{-5pt}{\bf .} \ } 
\newcommand{\ecl}{\end{cl}}
\newcommand{\bre}{\begin{re}\hspace{-5pt}{\bf .} \ }
\newcommand{\ere}{\end{re}}
\newcommand{\bas}{\begin{as}\hspace{-5pt}{\bf .} \ }
\newcommand{\eas}{\end{as}}
\newcommand{\bwg}{\begin{wg}\hspace{-5pt}{\bf .} \ }
\newcommand{\ewg}{\end{wg}}
\newcommand{\bex}{\begin{ex}\hspace{-5pt}{\bf .} \ }  
\newcommand{\eex}{\end{ex}}

\newcommand{\bit}{\begin{itemize}}
\newcommand{\eit}{\end{itemize}\par\noindent}
\newcommand{\beq}{\begin{equation}} 
\newcommand{\eeq}{\end{equation}\par\noindent}
\newcommand{\beqa}{\begin{eqnarray*}}
\newcommand{\eeqa}{\end{eqnarray*}\par\noindent}
\newcommand{\beqn}{\begin{eqnarray}}  
\newcommand{\eeqn}{\end{eqnarray}\par\noindent}

\newfont{\rmTen}{cmr10}

\newfam\msbfam
\font\tenmsb=msbm10				\textfont\msbfam=\tenmsb
\font\sevenmsb=msbm7			\scriptfont\msbfam=\sevenmsb
\font\fivemsb=msbm5				\scriptscriptfont\msbfam=\fivemsb

\begin{document}

\def\impll{\hbox{$\ -\hspace{-2mm}-\hspace{-1.6mm}\circ\ $}}
\def\simpll{\hbox{$\small\, -\hspace{-3mm}-\hspace{-0.5mm}\circ\, $}}
\def\retro{\hbox{$\small\, \circ\hspace{-1.5mm} -\hspace{-1mm}-\ $}}  
\def\sretro{\hbox{$\small\, \circ\hspace{-1.3mm} -\hspace{-2mm}-\ $}}

\centerline{\large{\bf A LOGICAL DESCRIPTION FOR PERFECT MEASUREMENTS}}
\par\bigskip\par   
\centerline{{\bf Bob Coecke} and {\bf Sonja Smets}\footnote{Department of Mathematics, and, 
Center for Logic and Philosophy of Science/Center Leo Apostel, Free University of Brussels, Pleinlaan
2, B-1050 Brussels, Belgium. E-mail: bocoecke@vub.ac.be and sonsmets@vub.ac.be. Bob Coecke is Post-Doctoral
Researcher and Sonja Smets Research Assistant, both at Flanders' Fund for Scientific Research. This
paper has been presented on IQSA's fourth biannual conference in Liptovski Jan, Slovakia, August
1998. We thank David Moore, Constantin Piron, Isar Stubbe and Frank
Valckenborgh for discussing matters related to the content of this paper.        
}}
\par
\bigskip
\par
\noindent
{\it Abstract:} 
We reconsider the description for property transitions
due to perfect measurements, viewing them as a special case of general transitions that are due to an
externally imposed change. We propose a corresponding syntax involving operational quantum logic and
a fragment of non-commutative linear logic.

\section{Introduction}     

In the spirit of Piron (1964, 1976, 1977),
Jauch and Piron (1969), Aerts (1982), Foulis {\it et al.} (1983), Foulis and Randall (1984), Faure {\it et al.}
(1995), Moore (1995, 1999), Amira {\it et al.} (1998) and Coecke and Stubbe (1999a, 1999b), we will point
out the correspondence between the act of `inducing properties'  (Amira {\it et al.}, 1998) and perfect
measurements (Piron, 1976) --- that is a joint ideal measurement of the first kind of a property and its
orthocomplement.  We describe the transitions that occur in such a perfect
measurement, and this will involve aspects of algebraic quantum logic (Birkhoff and von Neumann, 1936;
Husimi, 1937; Sasaki, 1954; Jauch, 1968; Piron, 1976; Kalmbach, 1983) and  
linear logic (Girard, 1987, 1989), in particular some of the non-commutative variants (Abrusci, 1990, 1991). 
Indeed, since general not-necessarily deterministic
property transitions yield quantale descriptions (Amira {\it et al.}, 1998; Coecke and Stubbe, 1999b) whereas
non-commutative linear logic yields quantale semantical models, the formal motivation for a logical description of
perfect measurements incorporating linear logical operations naturally arises.

In this paper, operational quantum logic (OQL) stands for the Geneva school approach to states and properties
of a physical entity ---we will not go into details and refer for the most recent
overview to Moore (1999).  The properties ${\cal
L}$ of the entity are ordered by an ---physically deducible--- implication relation, as such structured as a
poset which {\it proves} to be meet-complete as a semilattice (Piron, 1976), and as such also complete as a
lattice, the join given by  
$\vee_ia_i=\wedge\{a\in{\cal L}|\forall i:a\geq a_i\}$.  A property is said to be actual if it is {\it true}, i.e.,
any verification of it would yield a positive answer with certainty, and a state is then defined as the strongest
actual property an entity possesses. We call a set of properties an {\it actuality set} if at least one property in it
is actual. Note that if
$A$ is an {\it actuality set}, we have immediately that $\vee A:=\vee_{a\in A}a$ is an actual property, and we will
refer to
this `strongest property of which $A$ implies actuality' as a {\it definite actual property} 
of an actuality set.
 As shown by Piron (1964, 1976), a complete lattice yields a representation as the projection lattice
of a `generalized' Hilbert space if and only if it is atomistic, orthomodular and satisfies the covering law, as such
assuring a realization within standard quantum theory. In this paper it suffices to require  
besides completeness only orthomodularity of the lattice, the latter defined by: 
(i) there exists $^\perp:{\cal L}\to{\cal L}$
fulfilling
$a\leq b\Rightarrow b^\perp\leq a^\perp$,
$a\wedge a^\perp=0$, $a\vee a^\perp=1$, $a^{\perp\perp}=a$, and, 
(ii) $a\leq b$ implies $a\vee(a^\perp\wedge b)=b$.
In the second section we show how the scheme developed in Amira {\it et al.} (1998) and Coecke
and Stubbe (1999a, 1999b) `lifts' the Baer$^*$-semigroups considered as state transitions by Pool (1968)
---and introduced by Foulis (1960) as a natural collection of morphisms for orthomodular 
lattices that embeds the closed
orthogonal projections on this lattice--- to more general classes of morphisms
that express indeterministic transitions. In a third section we translate this in logical axioms, explicitly
expressing that these transitions are due to the interaction with an externally imposed context, and in the
particular case of quantum measurements a  `perfect measurement context'.    
  
\section{Propagation of properties due to a perfect measurement} 

The maps ${\cal P}({\cal L})=\{\varphi_a:{\cal L}\to{\cal L}:b\mapsto a\wedge(b\vee a^\perp)|a\in{\cal L}\}$
---the `Sasaki (1954) projections'--- prove to be the set of all closed orthogonal projections on ${\cal L}$
, and are by  
$\theta:{\cal L}\to{\cal P}({\cal L}):a\mapsto\varphi_a$ in isomorphic correspondence with ${\cal L}$ when
ordered by
$\varphi_a\unlhd \varphi_{a'}\Leftrightarrow\varphi_a\varphi_{a'}=\varphi_a$ (Foulis, 1960; Piron, 1995). However, the maps
$\varphi_a$ are not closed under their natural operation `composition', and should be considered as embedded in ${\cal
S}({\cal L}):=\{f:{\cal L}\to{\cal L}|f(\vee_ia_i)=\vee_if(a_i)\}$ the corresponding complete
Baer$^*$-semigroup: 
note here that ${\cal S}({\cal
L})$ is itself a join-complete lattice with respect to the pointwise computed order $f\leq f'\Leftrightarrow\forall
a\in{\cal L}:f(a)\leq f'(a)$, yielding a pointwise computable join for all
$\{f_i\}_i\subseteq{\cal S}({\cal L})$ as  
$\bigvee_i f_i:{\cal L}\to{\cal L}:a\mapsto\vee_if_i(a)$. However, the inclusion of ${\cal P}({\cal L})$ ---with
ordering inherited from  
${\cal L}$ through $\theta$--- in $({\cal S}({\cal L}),\bigvee)$ does not
preserve the partial order.  Indeed, for $a,a'\in{\cal L}$ with $a'\not\leq a\lood\not \leq 1$ and $a\wedge a'=0$ we
have
$a\leq a\vee a'$, but since $0<\varphi_a(a')\leq a$ and $\varphi_{a\vee a'}(a')=a'$ it follows that 
$\varphi_{a'}\wedge \varphi_{a\vee a'}(a') =0$ and thus 
$\varphi_a(a')\not\leq\varphi_{a\vee a'}(a')$ so that $\varphi_a\not\leq\varphi_{a\vee a'}$ although 
$\varphi_a\unlhd\varphi_{a\vee a'}$.  
Instead of considering the closed orthogonal projections ${\cal P}({\cal L})$, we will rather
consider 
${\cal P}^\#({\cal L})=\{\varphi_{\{a,a^\perp\}}|a\in{\cal L}\}$
with ---$P{\cal L}$ denotes the powerset of
${\cal L}\setminus\{0\}$:
\beq\label{eq:varphi}
\varphi_{\{a,a^\perp\}}:P{\cal L}\to P{\cal L}:
B\mapsto\{\varphi_a(b)|b\in B,b\not\leq a^\perp\}\cup\{\varphi_{a^\perp}(b)|b\in B,b\not\leq a\}
\eeq  
The maps in ${\cal P}^\#({\cal L})$ will be interpreted as describing the propagation of properties in perfect
measurements:
\par
\smallskip    
\par
\noindent
{\it
A property $b\in{\cal L}$ that is
actual before the measurement yields an actual property $\varphi_a(b)$ {\it or}\, $\varphi_{a^\perp}(b)$ 
after it.
In general neither $\varphi_a(b)$ nor $\varphi_{a^\perp}(b)$ `will be' actual with certainty after the
measurement, provided that $b \not \leq a$ and $b \not \leq a\lood$.  The strongest `definite actual property' for the
actuality set $\{\varphi_a(b),\varphi_{a\lood}(b)\}$ is $\varphi_a(b)\of\varphi_{a\lood}(b)$. 
It has been motivated that
 actual properties propagate preserving the
join (Pool, 1968; Faure {\it et al.}, 1995) as such giving the maps in ${\cal S}({\cal L})$ the
significance of describing propagation of actual properties. Therefore, given an initial actuality
set, we consider them as describing the propagations of the definite actual properties.}
\par
\smallskip  
\par
\noindent
It is
again natural to consider the maps $\varphi_{\{a,a\lood\}}$ as belonging to a more general collection closed under  
composition, but due to the formal change of domain from the
$\vee$-lattice
${\cal L}$ to the $\cup$-lattice $P{\cal L}$ this requires an essentially different
approach. 
A construction that allows this embedment is proposed in Amira {\it et
al.} (1998) and Coecke and Stubbe (1999a, 1999b) and will now be applied to the maps defined by
eq(\ref{eq:varphi}).
The set ${\cal P}^\#({\cal L})$ is canonically in surjective correspondence with ${\cal P}({\cal
L})$ and thus with ${\cal L}$ itself by $\phi:{\cal P}({\cal L})\to{\cal P}^\#({\cal L}):\varphi_{a}
\mapsto\varphi_{\{a,a^\perp\}}$ that factorizes in:  
\beq
\left\{  
\begin{array}{l}
\theta^P:{\cal P}({\cal L})\to{\cal P}^P({\cal L}):
\varphi_{a}\mapsto[P{\cal L}\to P{\cal L}:B\mapsto\{\varphi_a(b)|b\in B,b\not\leq a^\perp\}]  
\\
\eta:{\cal P}^P({\cal L})\to{\cal P}^\#({\cal L}): 
\theta^P(\varphi_a)\mapsto\theta^P(\varphi_a)\cup\theta^P(\varphi_{a^\perp})  
\end{array}
\right.
\eeq
where ${\cal P}^P({\cal L})$ is implicitly defined as the range of $\theta^P$. The following is
obvious.
\bpr\label{lm:123}  
(i) $\varphi_{\{a,a^\perp\}}=\varphi_{\{a^\perp,a^{\perp\perp}\}}$; 
(ii) $\phi(\varphi_a)=\phi(\varphi_b)\Leftrightarrow b\in\{a,a^\perp\}$.
\epr
Thus, ${\cal P}^\#({\cal L})$, ${\cal P}({\cal L})/\hspace{-1mm}\sim$ and ${\cal
L}/\hspace{-1mm}\sim$ are in bijective correspondence for the equivalence relation  
$\varphi_b\sim\varphi_a$ (respectively $a\sim b$) iff $b\in\{a,a^\perp\}$.
Clearly, the ordering of $({\cal P}({\cal L}),\unlhd)$ is in no way inherited by ${\cal
P}^\#({\cal L})$. However, as we will show next, the maps in ${\cal P}^\#({\cal L})$ can be considered
as incomparable with respect to the partial ordering inherited from $({\cal S}({\cal L}),\leq)$.
For maps $f:P{\cal L}\to P{\cal L}$, denote the condition
$\forall A,B\in P{\cal L}:\vee A=\vee B\Rightarrow\vee f(A)=\vee f(B)$ as $A^\#$, and 
define the following subset of ${\cal S}(P{\cal L})$:
\beq\label{Cross}
{\cal Q}^\#({\cal L}):=\{f:P{\cal L}\to P{\cal L}|f(\cup_iA_i)=\cup_if(A_i),f\ meets\ A^\#\}
\eeq
As shown in Amira {\it et
al.} (1998) and Coecke and Stubbe (1999a) it is exactly the condition $A^\#$ that forces the definite actual
properties to propagate preserving the join. We will now sketch how this connection between ${\cal
Q}^\#({\cal L})$ and
${\cal S}({\cal L})$ is realized. Set:
\beq 
\theta^P:{\cal S}({\cal L})\to{\cal S}(P{\cal L}):
f\mapsto[P{\cal L}\to P{\cal L}:B\mapsto\{f(b)|b\in B\}]          
\eeq 
Recall that a quantale (Mulvey, 1986) is a join-complete lattice equipped with an operation
$\circ$ that distributes at both sides 
over arbitrary joins. Quantale morphisms preserve $\circ$ and all joins.
The following proposition can be found in Coecke and Stubbe (1999a, 1999b).    
\bpr  
The set $P{\cal S}^P({\cal
L})=\bigl\{\bigcup_i\theta^P(f_i)\bigm|\forall i:f_i\in{\cal S}({\cal L})\bigr\}$ defines a strict subquantale
of
$({\cal Q}^\#({\cal L}),\bigcup,\circ)$, where $\bigcup$ denotes pointwisely computed unions and $\circ$
composition. Moreover, the map $\bigvee[-]:{\cal Q}^\#({\cal L})\to{\cal S}({\cal L}):f\mapsto[{\cal L}\to{\cal L}:a\mapsto\vee
f(\{a\})]$ is a surjective quantale morphism.
\epr
Clearly ${\cal P}^\#({\cal L})\hookrightarrow P{\cal S}^P({\cal L})$ which is closed under
composition, and since ${\cal P}^\#({\cal L})\hookrightarrow{\cal Q}^\#({\cal L})$, the corresponding definite actual
properties propagate preserving the join.
Indeed, given the union preserving map $\varphi_{\{a,a^\perp\}}$, the map describing the propagation of
definite actual properties is the join preserving map
$\bigvee[\varphi_{\{a,a^\perp\}}]:{\cal L}\to{\cal L}:b\mapsto\varphi_a(b)\vee\varphi_{a'}(b)$.
The following (non fully commutative!) scheme summarizes all the above situating ${\cal
P}({\cal L})$ and
${\cal S}({\cal L})$ relative to ${\cal P}^\#({\cal L})$ and ${\cal Q}^\#({\cal L})$:  
\beq\label{diagram}
\begin{array}{ccccccccc}  
{\cal L}&\stackrel{\cong}{\longleftrightarrow}&{\cal P}({\cal L})&\hookrightarrow&({\cal S}({\cal
L}),\bigvee,\circ)\\
\\
 & &\ \ \downarrow{\scriptstyle\phi}& & &\nwarrow&^{\bigvee[-]}\ \ \ \ \ \ \ \ \ \ \ \ \ \\
\\
 & & {\cal P}^\#({\cal L})&\hookrightarrow&({\cal P}{\cal S}^P({\cal L}),\bigcup,\circ)&\hookrightarrow&({\cal
Q}^\#({\cal L}),\bigcup,\circ)&
\hookrightarrow&({\cal S}(P{\cal L}),\bigcup,\circ)
\end{array}  
\eeq
We will now apply the above to the physical situation where a physical entity is placed in a measurement
context that induces a perfect measurement. Recall that $a,b\in{\cal L}$ are compatible if the Boolean
sublattice generated by
$\{a,a^\perp,b,b^\perp\}$ distributes. 
\bdf
An `induction' on a physical entity is an externally imposed change of properties.  A `perfect
measurement induction', characterized by a pair $\{a,a^\perp\}$, is an induction such that after
it the property $a$ is actual `or' the property $a^\perp$ is actual, and any property $b$ which is compatible
with
$a$ --- and as such also with $a^\perp$ --- that  is actual before is still actual afterwards.   
\edf
Using eq(\ref{eq:varphi}), Proposition \ref{lm:123} and Piron (1976) p.69 Theorem 4.3 one obtains:  
\bpr
Given a perfect measurement induction of $\{a,a^\perp\}$ on an entity with property lattice ${\cal L}$, the map
$\varphi_{\{a,a^\perp\}}:P{\cal L}\to P{\cal L}$ describes the propagation of actuality sets.
\epr
Clearly, these perfect measurement inductions can be interpreted as a minimal disturbance of the
entity assuring actuality of $a$ `or' $a^\perp$.  The essence of this remark boils down to the use of
`or' which in this case is a disjunction expressing what we could call inner non determinism:  
\par
\smallskip    
\par
\noindent 
{\it
Indeed,
crucial in the notion of a perfect measurement induction is that we express how two properties both
have an ability `to be' actual in case we are going to perform the induction.  This aspect
forces us to consider unions of transitions and underlying actuality sets, i.e., considering maps
in ${\cal Q}^\#({\cal L})$ rather than in ${\cal S}({\cal L})$. 
}  

\section{A logical description for perfect measurements}  

In this section we give logical axioms for the propagation of properties , more precisely of actuality sets, in
perfect measurements ---i.e., the maps above described by ${\cal P}^\#({\cal L})$ and their finite compositions---
explicitly expressing  that the measurement process is a non-deterministic transition due to interaction with an
externally imposed context ---see for example Gisin and Piron (1981).  A detailed description of the logical
language, containing the elementary and well formed formula's linked to physical entities, and sequent calculus for
describing property transitions in general is being elaborated on in Coecke and Smets (n.d.). In this paper we stick
to a combination of a fragment of non-commutative linear logic and OQL. We will need some of the Left and Right
sequent rules of non-commutative intuitionistic linear logic $NCIL$ as defined in Abrusci (1990), extended to a
predicate calculus:
\beqa
\begin{array}{cc}
(id.)\ :=\ {\over^{A\volgt A}}
\quad\quad &\quad\quad
(cut)\ :=\ {\Gamma\volgt A \quad \Gamma\1 ,A,\Gamma\2 \volgt\Delta \over \Gamma\1,\Gamma,\Gamma\2 \volgt\Delta}
\end{array}
\eeqa
\vspace{-7mm}
\beqa
\begin{array}{cc}
(\otimes,R)\ :=\ {\Gamma\1 \volgt A \quad \Gamma\2 \volgt B \over \Gamma\1,\Gamma\2 \volgt A \otimes B}
\quad\quad &\quad\quad
(\otimes,L)\ :=\ {\Gamma\1,A,B,\Gamma\2 \volgt \Delta \over \Gamma\1,(A \otimes B),\Gamma\2 \volgt\Delta}
\end{array}    
\eeqa
\vspace{-7mm}
\beqa
\begin{array}{ccc}
(\oplus,R1)\ :=\ {\Gamma \volgt A \over \Gamma \volgt (A \oplus B)} 
\quad\quad &\quad\quad
(\oplus,R2)\ :=\ {\Gamma \volgt B \over \Gamma \volgt (A\oplus B)} 
\quad\quad &\quad\quad
(\oplus,L)\ :=\  {\Gamma\1,A,\Gamma\2 \volgt \Delta \quad \Gamma\1,B,\Gamma\2 \volgt \Delta \over
\Gamma\1,(A\oplus B),\Gamma\2
\volgt \Delta }
\end{array}
\eeqa
\vspace{-7mm}
\beqa
\begin{array}{cc}
(\impll,R)\ :=\ {A,\GammaÊ\volgt B \over \Gamma \volgt AÊ\simpll B} 
\quad\quad &\quad\quad  
(\impll,L)\ :=\ {\Gamma\volgt A \quad \Gamma\1,B,\Gamma\2\volgt\DeltaÊ\over
\Gamma\1,\Gamma,(A\simpll B),\Gamma\2 \volgt\Delta}  
\end{array}
\eeqa
\vspace{-7mm}
\beqa
\begin{array}{cc}
(\forall ,R)\ :=\ {\Gamma\volgt A \over \Gamma\volgt \forall x A}\ (\ x\ not\ free\ in\
\Gamma )\
\quad\quad & \quad\quad
(\forall ,L)\ :=\ {\Gamma\1,A[y/x],\Gamma\2\volgt BÊ\over \Gamma\1,\forall x A,\Gamma\2 \volgt B}
\end{array}
\eeqa
We express the properties in ${\cal L}_r \setminus\{ 0 \}$ for physical entity $r$, as property-terms (constants and
variables) in our logical language.  We will limit ourselves to
the following primitive propositions:  
\beqa
In_r(x)&:=&x\ is\ actual\ for\ physical\ entity\ r\\ 
R_r(x)&:=&x\ and\ only\ x\ is\ reachable\ for\ physical\ entity\ r\\
M_r(x,x\lood)&:=&x\ or\ x\lood\ will\ be\ induced\ on\ physical\ entity\ r
\eeqa
where formula $M_r(x,x\lood)$
should be thought of as representing the measurement context imposed on the entity in order to induce
properties, and thus stands for the `induction' itself.

Given the fact that we work with OQL yielding an additional structure for our 
property-terms within the considered
fragment of non-commutative linear logic, we add the axioms which show essential 
OQL-features:
\beqa
\begin{array}{cc}
\otimes_{x \in X}[In_r(x)] \volgt In_r(\wedge X)  
\quad\quad &\quad\quad
In_r(x) \volgt In_r(x\of y)
\end{array}
\eeqa
All the following axioms express the content of an arbitrary
map $\varphi_{\{a,a^\perp\}}\in{\cal P}^\#({\cal L})$ by considering eq(\ref{eq:varphi}) and
incorporate explicitly the role of the induction context:
\beqa
\left\{
\begin{array}{c}
\underline{Trans}\ :\ 
\forall y \forall z : In_r(y) \otimes R_r(z) \impll In_r\bigl(\varphi_z(y)\bigr)\otimes R_r\bigl(\varphi_z(y)\bigr)\hfill
\vspace{2mm}\\
\underline{Adjust1}\ :\
\forall x \forall y\{\not\leq x,\not\leq x\lood\}: M_r(x,x\lood) \otimes [In_r(y) \otimes R_r(y)]  \impll In_r(y) \otimes
((R_r(x) \oplus R_r(x\lood))\hfill
\vspace{2mm}\\ 
\underline{Adjust2}\ :\
\forall x \forall y\leq x:  M_r(x,x\lood) \otimes [In_r(y) \otimes R_r(y)] \impll In_r(y) \otimes R_r(y)\hfill
\end{array}
\right.
\eeqa
The first axiom $\underline{Trans}$, expresses the induction of one reachable property according to the Sasaki
projection. The second axiom $\underline{Adjust1}$
expresses a re-adjustment of the entity relative to the imposed induction-context. With other words, it
expresses the act of imposing the induction-context
$M_r(x,x^{\perp})$ on the entity $r$ consisting of an actual and reachable property.  

Consider in the following application, an initial situation $ M(b,b\lood) \otimes [In(a) \otimes
R(a)]$ of a physical entity with actual property
$a$, assuming that $b \not \leq a$ and $b\not\leq a\lood$ ---we drop the
subscript $r$ since we consider only one physical entity.
Note that we use not only the expression of the current 
property which is actual for the entity but also the fact that this same property is reachable for the entity. 
It is now our aim to express the fact that the entity with property $a$ will end up with actually having one of the
possible properties $b$ or $b\lood$ which are reachable according to the given induction. 
\beq\label{LongEq}
\begin{array}{c}
M(b,b\lood) \otimes  [In(a) \otimes R(a)]  \volgt 
\bigl[In\bigl(\varphi_b(a)\bigr)
\otimes R\bigl(\varphi_b(a)\bigr)\bigr] \oplus \bigl[In\bigl(\varphi_{b^\perp}(a)\bigr)
\otimes R\bigl(\varphi_{b^\perp}(a)\bigr)\bigr]
\end{array}
\eeq
This is indeed an expression in our fragment of non-commutative linear logic and OQL which is provable
according to the   following subproofs which we can establish, using our axioms and logical tools:
\par\smallskip\par\noindent
1) We need to adjust our entity to the
given measurement context (update $R$, given $M$), using $\underline{Adjust1}$. We need the elimination of the universal
quantifier twice,  and Modus Ponens `$A,A\simpll B\volgt B$':
\beq 
M(b,b\lood) \otimes  [In(a) \otimes R(a)]  \volgt In(a) \otimes (R(b) \oplus
R(b\lood))
\eeq 
2) Using distributivity of $\otimes$ over $\oplus$\footnote{See in Abrusci(1990) the fragment of
definition 2.1 :``...$\langle
X,\leq,1,\perp,\top,\otimes,\en,\oplus,\hspace{-1mm}\simpll,\hspace{-1mm}\sretro\rangle$ is a non-commutative
intuitionistic *-linear structure iff ... (xvii) $\forall x \in X \forall y \in X \forall z \in X: z \otimes (x
\oplus y) =(z
\otimes x)\oplus(z\otimes y)$.''}: 
\beq
In(a) \otimes (R(b) \oplus R(b\lood) ) \volgt [In(a) \otimes R(b)] \oplus    
[In(a)
\otimes R(b\lood)]
\eeq
3) Using the axiom $\underline{Trans}$, the elimination of
the universal quantifier twice, and Modus Ponens:
\beq
In(a) \otimes R(b) \volgt In(\varphi_{b}(a)) \otimes R(\varphi_{b}(a))
\eeq
4) Using the axiom $\underline{Trans}$, the elimination of the universal quantifier  
twice, and Modus Ponens, then:
\beq
In(a) \otimes R(b\lood) \volgt In(\varphi_{b\lood}(a)) \otimes R(\varphi_{b\lood}(a))
\eeq
Finally by the previous subproofs we obtain our goal, {\it i.e.},
eq(\ref{LongEq}).  
\par
\medskip
\par
\noindent
{\it Some additional remarks:}  
\par
\medskip
\par
\noindent
(i) In our formalism we can perform a succession of inductions. Referring back to
eq(\ref{eq:varphi}),
we can rediscover $\varphi_{\{c,c^\perp\}}\varphi_{\{b,b^\perp\}}(\{a\}) = 
\{\varphi_{c}(\varphi_{b}(a)),
\varphi_{c\lood}(\varphi_{b}(a)), \varphi_{c}(\varphi_{b\lood}(a)),
\varphi_{c\lood}(\varphi_{b\lood}(a))\}$ as ---for those $c$ that satisfy it: 
\beqa
\begin{array}{l}
M(c,c\lood) \otimes \lgroup [In(\varphi_{b}(a))  \otimes R(\varphi_{b}(a))] \oplus 
[In(\varphi_{b\lood}(a)) \otimes R(\varphi_{b\lood}(a))]\rgroup \volgt\vspace{2mm} \\ 
\hspace{4.3cm}{[}In(\varphi_{c}(\varphi_{b}(a))) \otimes R(\varphi_{c}(\varphi_{b}(a)))] \oplus 
[In(\varphi_{c\lood}(\varphi_{b}(a))) \otimes R(\varphi_{c\lood}(\varphi_{b}(a)))]\vspace{1mm}  \\ 
\hspace{4.3cm}\oplus{[}In(\varphi_{c}(\varphi_{b\lood}(a))) \otimes R(\varphi_{c}(\varphi_{b\lood}(a)))] \oplus 
[In(\varphi_{c\lood}(\varphi_{b\lood}(a))) \otimes R(\varphi_{c\lood}(\varphi_{b\lood}(a)))]  
\end{array}
\eeqa 
Remark that $M(c,c\lood) \otimes \lgroup [In(\varphi_{b}(a)) \otimes R(\varphi_{b}(a))]
\oplus [In(\varphi_{b\lood}(a))
\otimes R(\varphi_{b\lood}(a))]\rgroup$ could also have been expressed by $M(c,c\lood) \otimes \lgroup
M(b,b\lood) \otimes [In(a) \otimes R(a)] \rgroup$, which points at the fact that we have to restrict 
associativity for the multiplicative conjunction. 
\par
\medskip  
\par
\noindent
(ii) This logical description generalizes to all transitions of properties    
described in
${\cal Q}^\#({\cal L})$. Set: 
\beqa
\begin{array}{c}
IND_r(\alpha)\ := The\ physical\ entity\ r\ is\ placed\ within\ the\  
context\ inducing\ propagation\ \alpha
\end{array}
\eeqa
\par\noindent
where $\alpha$ takes values $f:P{\cal L}\to P{\cal L}$ in ${\cal Q}^\#({\cal L})$ and set
$K_f=\{a\in{\cal L}|f(\{a\})=\emptyset\}$. The following axiom expresses the content of the maps in ${\cal Q}^\#({\cal L})$:
\beqa
\underline{General\ Propagation}\ : \forall \alpha \forall x\not\in K_\alpha:  IND_r(\alpha) \otimes In_r(x)
\impll
\oplus _{z \in \alpha(\{x\})}In_r(z)
\eeqa
yielding formal implementation of $A^\#$ as $IND_r(\alpha) \otimes In_r(\vee X) \impll In_r(\of \alpha
(X))$ implicitly for $\alpha(X) \not = \emptyset$.
We recover the combined axioms $\underline{Trans}$,
$\underline{Adjust1}$ and
$\underline{Adjust2}$ for a particular $\varphi_{\{y,y\lood\}}$ as:
\beqa
\begin{array}{c}
\forall x : IND_r(\varphi_{\{y,y\lood\}}) \otimes In_r(x) \impll
\oplus _{z \in \varphi_{\{y,y\lood\}}(\{x\})} In_r(z)\hfill
\vspace{2mm}\\
\end{array}
\eeqa

\section{References}    
{
\noindent 
Birkhoff, G., and von Neumann, J. (1936) {\em Ann. Math.} {\bf 37}, 823.  
\par\vspace{0cm}\par      
\noindent 
Husimi, K. (1937) {\it Proc. Physicomath. Soc. Japan} {\bf 19}, 766
\par\vspace{0cm}\par      
\noindent 
Sasaki, U. (1954) {\it J. Sci. Hiroshima Univ.} {\bf A17}, 293.
\par\vspace{0cm}\par      
\noindent 
Foulis, D.J. (1960) {\it Proc. AMS} {\bf 11}, 648.  
\par\vspace{0cm}\par      
\noindent 
Jauch, J.M. (1968) {\it Foundations of Quantum Mechanics}, Addison-Wesley.
\par\vspace{0cm}\par      
\noindent 
Jauch, J.M., and Piron, C. (1969) {\it Helv. Phys. Acta} {\bf 42}, 842.
\par\vspace{0cm}\par      
\noindent 
Piron, C. (1964) {\it Helv. Phys. Acta} {\bf 37}, 439.
\par\vspace{0cm}\par      
\noindent 
Pool, J.C.T. (1968) {\it Comm. Math. Phys.} {\bf 9}, 118.
\par\vspace{0cm}\par        
\noindent 
Piron, C. (1976) {\it Foundations of Quantum Physics}, W. A. Benjamin, Inc.
\par\vspace{0cm}\par      
\noindent 
Piron, C. (1977) {\it J. Phyl. Logic} {\bf 6}, 481.
\par\vspace{0cm}\par      
\noindent 
Gisin, N. and Piron, C. (1981) {\it Lett. Math. Phys.} {\bf 5}, 379.
\par\vspace{0cm}\par      
\noindent 
Aerts, D. (1982) {\it Found. Phys.} {\bf 12}, 1131.
\par\vspace{0cm}\par        
\noindent 
Foulis, D.J., Randall, C.H., and Piron, C. (1983) {\it Found. Phys.} {\bf 13}, 813.
\par\vspace{0cm}\par      
\noindent 
Foulis, D.J. and Randall (1984) {\it Found. Phys.} {\bf 14}, 65.
\par\vspace{0cm}\par      
\noindent 
Kalmbach, G. (1983) {\it Orthomodular Lattices}, Academic Press.
\par\vspace{0cm}\par      
\noindent 
Mulvey, C.J. (1986) {\em Rend. Circ. Math. Palermo} {\bf 12}, 99.
\par\vspace{0cm}\par      
\noindent 
Girard, J.Y. (1987) {\it Theor. Comp. Sc.} {\bf 50}, 1.  
\par\vspace{0cm}\par      
\noindent 
Girard, J.Y. (1989) {\it Contemporary Mathematics} {\bf 92}, 69.  
\par\vspace{0cm}\par      
\noindent 
Abrusci, V.M. (1990) {\it Zeitschr. Math. Logik Grundl. Math.}, {\bf 36}, 297.
\par\vspace{0cm}\par      
\noindent 
Abrusci, V.M. (1991) {\it The journal of Symbolic Logic} {\bf 56}, 1403.   
\par\vspace{0cm}\par      
\noindent 
Faure, Cl.-A., Moore, D.J., and Piron, C. (1995) {\it Helv. Phys. Acta} {\bf 68}, 150.  
\par\vspace{0cm}\par      
\noindent 
Moore, D.J. (1995) {\it Helv. Phys. Acta} {\bf 68}, 658.
\par\vspace{0cm}\par      
\noindent 
Piron, C. (1995) {\it Int. J. Theor. Phys.} {\bf 34}, 1681.
\par\vspace{0cm}\par      
\noindent 
Amira, H., Coecke, B., and Stubbe, I. (1998) {\it Helv. Phys. Acta} {\bf 71}, 554.  
\par\vspace{0cm}\par      
\noindent 
Coecke, B., and Stubbe, I. (1999a) {\it Found. Phys. Lett} {\bf 12}, 29.
\par\vspace{0cm}\par      
\noindent 
Coecke, B., and Stubbe, I. (1999b) `State Transitions as Morphisms for Join
Complete lattices', {\it Int. J. Theor. Phys.} {\bf 39}, 605.
\par\vspace{0cm}\par
\noindent
Moore, D.J. (1999) `On State Spaces and Property Lattices', {\it Stud$.$ Hist$.$
Phil$.$ Mod$.$ Phys.} {\bf 30}, 61.  
\par\vspace{0cm}\par      
\noindent 
Coecke, B., and Smets, S. (n.d.) `Combining Quantum and Linear Logic for Propagating Physical
Properties', in preparation. 
\noindent}
\end{document}